     \documentstyle[12pt]{article}

\renewcommand{\theequation}{\arabic{section}.\arabic{equation}}

\renewcommand{\thesection}{\arabic{section})}

     \newcommand{\be}{\begin{equation}}
     \newcommand{\ee}{\end{equation}}
     \newcommand{\ba}{\begin{eqnarray}}
     \newcommand{\ea}{\end{eqnarray}}
     \newcommand{\spa}{\hspace{0.5cm}}

     \newfont{\myfont}{msbm10 scaled\magstep1}
     \newcommand{\del}{\partial}

\newcommand{\cf}{{\cal F}}

\newcommand{\cl}{{\cal L}}

\newcommand{\fr}{\frac}

\begin{document}

      \begin{titlepage}

\topmargin -1cm

\rightline{PUPT-1404}
\rightline{June 1993}

     \vskip 10mm

 \centerline{ \LARGE\bf Are There Topologically Charged States }
 \vskip 3pt
\centerline{ \LARGE\bf  Associated with Quantum Electrodynamics ? }

    \vskip 2.0cm

    \centerline{\sc E.C.Marino \footnote{On sabbatical leave from
Departamento de F\'{\i}sica, Pontif\'{\i}cia Universidade
Cat\'{o}lica, Rio de Janeiro, Brazil. E-mail :
marino@puhep1.princeton.edu }}

     \vskip 0.6cm

     \centerline{\it Joseph Henry Laboratories }
    \centerline{\it   Princeton University }
    \centerline{\it    Princeton, NJ 08544 }
\vskip 1.0cm

     \vskip 1.0cm

  \begin{abstract}
 We present a formulation of
 Quantum Electrodynamics in terms of
  an antisymmetric tensor gauge field. In this
formulation the topological current of this field appears as a
 source for the electromagnetic field and the topological
charge therefore acts physically as an electric charge. The charged
 states of QED lie in the sector where the topological charge
is identical to the matter charge.
 The antisymmetric field theory, however,
admits new sectors where the topological charge is more general.
 These nontrivial, electrically charged, sectors contain massless
states orthogonal to the vacuum which are created by a gauge invariant
operator and can be interpreted as coherent states of photons.
We evaluate the correlation functions of these states
 in the
absence of matter. The new states have a positive definite
norm and do interact with the charged states of QED in the usual way.
It is argued that if these new sectors are in fact realized
in nature then a very intense background electromagnetic
field is necessary for the experimental observation of them.
 The order of magnitude of the intensity threshold is estimated.

\end{abstract}

\vfill
\end{titlepage}

\hoffset= -4mm

\leftmargin 25mm

\topmargin -8mm
\hsize 153mm

\baselineskip 7mm
\setcounter{page}{2}

\section{Introduction}

  \spa An extremely interesting feature of many quantum field theories
is the presence of topological sectors in the spectrum of
quantum excitations. These, to use a broad definition, are
characterized by the fact that they carry a charge which is
associated to an identically conserved current rather than to one
 whose conservation stems from a global
symmetry of the lagrangian, via Noether theorem. A general feature
of this kind of excitations is the fact that their
physical behavior cannot be decribed in terms of the lagrangian
fields or even in terms of polynomials in these fields. This
fact immediately rules out the possibility of any perturbative
treatment of them.

 There is a large diversity of systems in various number of dimensions
displaying quantum topological excitations in their spectrum. Usually
they are associated to finite energy classical solutions of the
field equations, known generally as solitons, of which they are the
quantum counterparts. In this case, the quantum solitons are
 massive, the quantum mass being of the order of the total energy
of the classical solution. Well known examples are the kinks,
Sine-Gordon solitons, vortices and magnetic monopoles \cite{col,evora}
 There are, however, less known examples where the system possesses
quantum topological excitations even though there are no corresponding
classical finite energy solutions of the field equations. These
occur, for example, in Z(4) symmetric systems of field theory or
quantum statistical mechanics in two dimensional spacetime,
whenever there are phases where the symmetry is partially broken
down to Z(2) \cite{evora,kob}.

 Quantum Electrodynamics in 3+1 dimensions is a theory which does
not possess either classical or quantum topological excitations.
 In this work, however, we show that QED can be formulated as
a particular sector of a theory for the antisymmetric tensor
gauge field $W_{\mu\nu}$. This field possesses a topological current
which appears as source for the electromagnetic field.
The topological charge, as a consequence, acts physically as an electric
charge. The usual (matter) charged states of QED lie in the sector
where the topological charge is identical to the matter charge.
 The theory, however, possesses new nontrivial topologically
charged sectors which do not belong to QED. The charge of this
states does not have its origin in matter. It is a state of the
gauge field.

 We construct a gauge invariant operator which creates the
topologically charged excitations in the new sectors and compute
its correlation functions in the absence of charged matter. These
can be taken to describe the large distance
regime of the correlation functions
when charged matter is present.
 The long distance behavior of the correlation functions indicates
that the new topologically charged states are massless and
orthogonal to the vacuum. From the two-point function, one can
also infer that the norm of these states is strictly positive
definite.
 The new toplogically charged states interact among themselves
as well as with the usual matter charged states in the
usual way as described by QED. When we express their creation
operator in terms of the electromagnetic field we see that
they can be interpreted as a coherent state of photons.

 There is a two-dimensional system which presents a very
similar structure as our four-dimensional theory for the
antisymmetric tensor gauge field. This is the free real massless
scalar field in 1+1 D. The main features of this theory
are reviewed in Section 2. An important difference of the
system considered here from the two-dimensional analog is the
fact that for a certain value of the topological charge
there are asymptotic states interpolated by the creation
operator of topological excitations. This fact allows us to
determine the value of the topological charge quantum.

 It may well happen that the more general formulation of QED
which admits these new charged sectors is simply not realized
in nature. If, on the other hand, the contrary is true, we argue
that an extremely strong background electromagnetic would
be required for the observation of the new states. We make
an estimate of the order of magnitude of the threshold magnitude
of the intensity of this background field.

 In Section 3, we present the formulation of QED in terms of the
antisymmetric tensor gauge field. In Section 4, we introduce the
creation operator of the quantum topological excitations. In
Section 5 we compute the correlation functions of this operator.
  Some phenomenological considerations concerning
the possibility of observation of the topological
states are made in Section 6. Conclusions and questions are
presented in Section 8. Four Appendixes are included in order
to demonstrate useful results.

\section{An Analogous System in 1+1 D}

\setcounter{equation}{0}

Let us review in this section an extremely simple system which
presents essentially the same features as the antisymmetric
tensor gauge field theory we found to be associated to QED in
3+1 D. This is the real massless scalar field in 1+1 D :
\be
\cl =\del_\mu\phi \del^\mu \phi
\label{2.1}
\ee
This field possesses an identically conserved topological current
\be
J^\mu = \epsilon^{\mu\nu}\del_\nu \phi
\label{2.2}
\ee
In terms of it the field equation $\Box \phi =0$ can be written as
\be
\epsilon^{\mu\nu} \del_\mu J_\nu =0
\label{2.3}
\ee
This admits the operator solution
\be
J^\mu =\del^\mu \chi \  \Longrightarrow \  \Box \chi=0
\label{2.4}
\ee

No polynomial in $\phi$ or its derivatives can create states
carrying the topological charge $Q =\int dx J^0$. Nevertheless,
if we introduce the operator
\be
\sigma (x,t) = \exp \left \{ ib \int^x \dot\phi(z,t) dz \right \}
 \  \Rightarrow \
\sigma(x,t) = \exp \left \{ i b \chi (x,t) \right \}
\label{2.5}
\ee
it is easy to see, by using canonical commutation relations,
that
\be
[Q,\sigma(x)] = b \sigma(x)
\label{2.6}
\ee
 This shows that $\sigma$ creates states bearing b units of
the topological charge $Q$ in spite
of the fact that the theory possesses no
classical soliton solutions. In the above expressions, $b$
is an arbitrary real parameter.

The topologically charged states will have  nontrivial
interactions even though the original theory is free. We
can figure out what this interaction will be by simply
writing the original lagrangian       in terms of $J^\mu$:
\be
{\cal L} =(1/2) \partial_\mu
\phi\partial^\mu\phi =(1/2) \epsilon^{\mu\alpha}\partial_\alpha \phi
\epsilon_{\mu\beta}\partial^\beta \phi    \  \Rightarrow \
{\cal L} = (1/2) J^\mu J_\mu
\label{2.7}
\ee
This indicates that the topological states will have a Thirring-like
interaction which is known to be the correct one.

We can easily compute the $\sigma$ euclidean correlation functions to be
\be
<\sigma(x)\sigma^\dagger(y)>= |x-y|^{- \fr{b^2}{8\pi}}
\label{2.8}
\ee
The long distance behavior of this implies that $<\sigma>=0$
and therefore the sigma operator creates nontrivial states
orhtogonal to the vacuum. Going to euclidean space and
taking the limit $x\rightarrow y$ we can see that the norm
of the $\sigma$ states is strictly positive, in spite of the
fact that the quantization of the free massless scalar field
in to dimensional spacetime requires a Hilbert space with an
indefinite metric \cite{W}. We are going to see that the
theory of the antisymmetric tensor gauge field we are going to
introduce in 3+1 D in connection to QED presents a structure
quite similar to the one just described.

\section{ QED Formulated as a Theory for the Antisymmetric
Tensor Gauge Field }

\setcounter{equation}{0}

Let us show in this section that QED can be formulated as
a particular sector of a theory for the antisymmetric
(Kalb-Ramond) tensor gauge field.
Let us start by the case in which matter is absent. Consider
the following theory for the antisymmetric tensor gauge field
$W_{\mu\nu}$:
\be
  {\cal L}_{W}
 = -\fr{1}{12} W_{\mu \nu \alpha}(-\Box)^{-1}W^{\mu \nu
\alpha}
\label{lw}
 \ee
where $ W_{\mu \nu \alpha} = \partial _{\mu}W_{\nu \alpha}+
\partial _{\nu}W_{\alpha \mu}+\partial _{\alpha}W_{\mu \nu}$ is the
field intensity tensor of the antisymmetric field.
The kernel in (\ref{lw}) is
\be
(-\Box^{-1}) = \int \fr{d^4 k}{(2\pi)^4} \fr{e^{i k\cdot (z-z')}}
{k^2} =\fr{1}{8 \pi^2 |z-z'|^2}
\label{ker}
\ee
All over this paper,
 of course, multiplication by $\Box ^{-1}$ is understood in
the convolution sense. The topological current of the
$W_{\mu\nu}$ field is given by
\be
J^\mu=
 \frac{1}{2}\epsilon^{\mu\nu
\alpha\beta}\partial_\nu W_{\alpha\beta}
\label{J}
\ee
 The field equation associated to (\ref{lw}) is
\be
\Box^{-1}\partial_\alpha W^{\alpha\mu\nu}=0
\label{fe}
\ee
We can rewrite this in terms of the topological current as
\be
\Box^{-1} \epsilon^{\mu\nu\alpha\beta}\partial_\alpha J_\beta=0
\label{weq}
\ee
We immediately see that we can obtain an operator solution for
$J^\mu$ in terms of a free masless scalar field
\be
J^\mu = \del ^\mu \chi \ \ \ \ \ \ \Box \chi =0
\label{jotachi}
\ee
Since $J^\mu$ has dimension 3, it follows that $\chi$ is a noncanonical
massless free field. We can express $J^\mu$ in terms of a
canonical masless free field  $\phi$ by using the pseudodifferential
operator $(-\Box)^{1/2}$:
\be
J^\mu = \partial^\mu (-\Box)^{1/2}\phi
\label{jotaphi}
\ee
Using $\Box \phi=0$, it is clear that
clear that
\be
\partial_\mu J^\mu=\Box(-\Box)^{1/2}\phi =(-\Box)^{1/2}\Box
\phi=0
\label{dej}
\ee

 As in our two-dimensional example,
no polynomial in the field intensity $W_{\mu\nu\alpha}$
can create states carrying the topological charge
$Q=\int d^3x J^0$. Nevertheless,
 we are going to see in the next section that it is possible to
construct an operator, analogous to the $\sigma$- operator of the
last section, which is going to create the topologically charged
states.

 As before, let us try to see what kind of interaction the
topological charge carrying states will have. In order to
do this, let us express the $\cl_W$ lagrangian in terms of the
topological current. Using (\ref{lw}) and (\ref{J}) we get
\be
{\cal L}_W= \frac{1}{2} J^\mu(-\Box)^{-1}J_\mu
\label{lj}
\ee
which is exactly the effective electromagnetic interaction between
charged particles associated with a current $J^\mu$
which would be obtained in QED upon integration over the
elctromagnetic field. For static
point chrges, for instance, the energy corresponding to (\ref{lj})
is the 1/r  Coulomb potential energy. We see that the topological
charges present an electromagnetic-like interaction among
themselves. This fact justifies our peculiar form of the
lagrangian $\cl_W$ containing the kernel $\Box^{-1}$.

 The fact that we can
describe the properties of QED through a theory like the one
given by ${\cal L}_W$, however, only becomes evident when we couple
the $W_{\mu\nu}$- field with the charged matter
current $j^\mu$ in the following
way:

\be
{\cal L}_W +{\cal L}_I =
  -\fr{1}{12} W_{\mu \nu \alpha}(-\Box)^{-1}W^{\mu \nu
\alpha} -\frac{q}{2} \epsilon^{\mu\nu\alpha\beta}
\partial_\nu W_{\alpha\beta} (-\Box)^{-1}j_\mu
\label{ltot}
\ee
where $q$ is the charge coupling constant of matter.

 In order to get the effective interaction of the matter
current generated by (\ref{ltot}), let us consider the
euclidean functional integration over $W_{\mu\nu}$ in
(\ref{ltot}):
$$
Z[j^\mu]=
  Z^{-1}\int DW_{\mu\nu} \exp\{-\int d^4z[ {\cal L}_W +{\cal L}_I
+{\cal L}_{GFW}]\}=
$$
\be
\exp \left \{-\fr{q^2}{2} \int d^4z d^4z' \left [(-\Box) j^\mu
\epsilon^{\mu\nu\alpha\beta} \del_\nu \right ]
\left [ (-\Box) j^\sigma \epsilon^{\sigma\lambda\gamma\rho}
 \del '_\lambda \right ] D^{\alpha\beta\gamma\rho}(z-z')\right \}
\label{eff1}
\ee

In the above expression ${\cal L}_{GFW}$ is the gauge fixing term,
given by
\be
{\cal L}_{GFW}=-(\xi /8)W_{\mu \nu}K^{\mu \nu \alpha \beta}
(-\Box)^{-1}W_{\alpha \beta}
\label{gf}
\ee
where $K^{\mu \nu \alpha \beta}=\partial^{\mu}
\partial^{\alpha}\delta^{\nu \beta}+\partial^{\nu}\partial^{\beta}
\delta^{\mu \alpha}-(\alpha \leftrightarrow \beta )$ and $\xi$ is
gauge fixing parameter and
$D^{\mu\nu\alpha\beta}$ is the euclidean propagator of the
 $W_{\mu\nu}$ field:
\be
D^{\mu\nu\alpha\beta}(x)=(1/4)[(-\Box)\Delta^{\mu\nu\alpha\beta}+
(1-\xi^{-1})K^{\mu\nu\alpha\beta}][8\pi^2\ |x|^2]^{-1}
\label{pro}
\ee
where  $\Delta^{\mu\nu\alpha\beta}=\delta^{\mu\alpha}\delta
^{\nu\beta}-\delta^{\mu\beta}\delta^{\nu\alpha}$.
Inserting (\ref{pro}) in (\ref{eff1}) we see that
 only
the first term of (\ref{pro}) contributes to (\ref{eff1}). Using
the fact that $j^\mu$ is a conserved current, we get
\be
Z[j^\mu] =
 \exp\{\frac{q^2}{2}\int d^4zd^4z' j^\mu(z)
(-\Box)^{-1} j_\mu(z')\}
\label{eff}
\ee

 The effective interaction $Z[j^\mu]$, given
by (~\ref{eff}) is precisely the one
which is generated by Maxwell QED upon integration over the
electromagnetic field. We therefore conclude that the theory
given by (~\ref{ltot}) does describe the same interaction of the
charged matter coupled to the antisymmetric field as does Maxwell QED.

 The field equation associated with (~\ref{ltot})  is
\be
\partial_\alpha W^{\mu\nu\alpha}=
 q \epsilon^{\mu\nu\alpha\beta}
\partial_\alpha j_\beta
\label{jeq}
\ee
 We can reexpress this in terms of the topological current as
\be
\epsilon^{\mu\nu\alpha\beta}\partial_\alpha
J_\beta= q \epsilon^{\mu\nu\alpha\beta}\del_\alpha j_\beta
\label{jeq1}
\ee
by using the identity
\be
\partial_\alpha W^{\mu\nu\alpha}\equiv
\epsilon^{\mu\nu\alpha\beta}\partial_\alpha
J_\beta
\label{id}
\ee

 From (\ref{jeq1}) it is clear that we can obtain an operator solution
for
 $J^\mu$ in terms of the matter
current and a free massless scalar field:
\be
J^\mu= qj^\mu +\partial^\mu \chi
\label{j}
\ee
where $\Box\chi=0$. As before, we could express $J^\mu$ as well in terms of
a canonical free massless field.

 Let us express now the $W_{\mu\nu}$ field directly in terms of the
electromagnetic field.
 We see that ${\cal L}_I$  can be
put exactly in the form of a minimal coupling of $j^\mu$ with the
electromagnetic field $A_\mu$, by defining (in the Lorentz gauge,
$\partial_\mu A^\mu=0$)
\be
J^\mu \equiv \partial_\nu F^{\nu\mu}
\label{wa}
\ee
 This can also be written in the equivalent form
\be
W^{\alpha\mu\nu} \equiv \epsilon^{\mu\nu\alpha\beta}(\Box)A_\beta
\label{wa1}
\ee
 The equivalence between (\ref{wa}) and (\ref{wa1}) is another form
 of the identity (\ref{id}).
 We could also obtain  (\ref{wa}) and (\ref{wa1}) as field equations
provided we make the substitution
\be
(\Box)^{-1}j^\mu \equiv A^\mu
\label{ja}
\ee
in $\cl_I$ , eq. (\ref{ltot}), obtaining thereby
\be
\cl_I = \epsilon^{\mu\nu\alpha\beta}A_\mu \del_\nu W_{\alpha\beta}
\label{li}
\ee
 Observe that (\ref{ja}) is precisely the relation between
$j^\mu$ and $A_\mu$ one would obtain from QED in the Lorentz
gauge.

 If we substitute
 (~\ref{wa}) in (~\ref{jeq1}) we get
\be
\epsilon^{\mu\nu\alpha\beta}\partial_\alpha[\partial^\gamma
F_{\gamma\beta}] = q \epsilon^{\mu\nu\alpha\beta}\partial_\alpha
[j_\beta]
\label{newmax}
\ee
which can be thought of as a generalized form of the Maxwell
equations. The general operator solution of (\ref{newmax})
is
\be
\partial_\nu F^{\nu\mu} = q j^\mu + \partial^\mu \chi
\label{newmax1}
\ee
 where $\Box \chi =0$. We could also have obtained (\ref{newmax1})
by inserting (\ref{wa}) in (\ref{j}).

 We see
from (~\ref{newmax})  that (~\ref{ltot}) provides a
 description of the electromagnetic interaction  which is
 more general than the one of QED because it admits solutions of the type
(~\ref{newmax1}) which contain the additional $\chi$ dependent
term.
 The usual states of QED lie in
the sector where $J^\mu =q j^\mu$, that is, the vacuum sector
of the free scalar field $\chi$. In the $W_{\mu\nu}$- field
formulation, however, we have new nontrivial sectors
for which the topological charge is nonzero even in the absence
 of matter, for instance.
 For these states,
the charge associated with
$J^\mu=\partial^\mu\chi$  would be nonzero.
 The topological current of the $W_{\mu\nu}$-field appears as a
source for the electromagnetic field. Since the solution for
$J^\mu$, eq. (\ref{j}) contains a piece which is independent of
the matter charge, we see that this formulation of QED admits
new charged sectors which are not related to matter.

 We can ask ourselves at this point about the interaction
between these new topologically charged states and the usual
charged matter states of QED. Observing that the interaction
lagrangian can be written as
\be
{\cal L}_I = q
J^\mu(-\Box)^{-1}j_\mu= J^\mu A_\mu
\label{qint}
\ee
we conclude that the topological states would interact
with the charged states of QED in the usual way. Also,
they would emit electromagnetic radiation in the same way as
the charged states of QED.

  The quantization of theories having nonlocalities of the
type appearing in (\ref{lw}) has been studied in detail
in \cite{ama} for
an arbitrary power of $-\Box$. There it is shown that these theories are
well behaved and sensible. The particular lagrangian
$\cl_W$ considered here,
where the power is -1 is actually local since it is a mass term for the
transverse part of $W_{\mu\nu}$.

 In the next section we are going to show that despite the
absence of finite energy solutions carrying the topological charge
we can construct an operator, analog to the $\sigma $-operator
of our two-dimensional example, which is going to create the
states in the new nontrivial toplogical sectors.

\section{ The $\mu$ Operator }

\setcounter{equation}{0}

 In this section we are going to introduce the operator which
will create the states in the nontrivial toplogical sectors of the
$W_{\mu\nu}$-field theory.
 This belongs to a class of operators which
was introduced in two, three, and four
dimensional spacetime as the creation operators of the respective
topological excitations, namely$-$ kinks, vortices and magnetic monopoles
\cite{kinks,vortices,mon,emprinc}.
 The method of construction of these operators
relies on the fact that the operator which creates the topological
excitations of a certain theory must behave as a disorder variable
and therefore be dual to the basic lagrangian fields.
As a consequence, these operators must satisfy a certain order-
disorder or dual algebra with the lagrangian fields \cite{evora}.

 A basic ingredient in the construction of topological
excitations creation operators is always \cite{evora}
a singular external field which, when coupled to the dynamical
fields produces the desired operator. In the present case,
the external field is the antisymmetric tensor
\be
\tilde{A} _{\mu \nu}(z,x)=\fr{b}{4\pi}\int _{T_{x}(S)}d^{3}\xi _{\mu}
\Phi _{\nu}(\xi-x)\delta ^{4}(z-\xi)  -(\mu \leftrightarrow \nu)
\label{ext}
\ee
where b is an arbitrary dimensionless real parameter and $\Phi _{\nu}=
(0,0,0,\frac{1-\cos\theta}{r\sin\theta})$ (in the coordinate system
$(t,r,\theta,\varphi)$). The integral in (~\ref{ext}) is performed over
the three-dimensional hypersurface $T_{x}(S)$ whose surface element
$d^{3}\xi _{\mu}$ has only the 0-component nonvanishing. $T_{x}(S)$ is the
region of the {\myfont R}$^{3}$ space external to the surface S at
$z^{0}=x^{0}$. This
surface is depicted in Fig. 1 and
 consists of a piece of sphere  of radius $\rho$
centered at $\vec{x}$ ($0\leq \theta
\leq \pi -\delta$) superimposed to an infinite trunk of
 cone with vertex at $\vec{x}$ and angle $\delta$
(cut a distance $\rho \cos \delta$ from the tip)
 with axis along $\theta =\pi $.
 In the region $T_x(S)$, the vector $\vec \Phi$ satisfies
the following useful identity
\be
\vec{\nabla}\times \vec{\Phi}\equiv\vec{\nabla} \left [-\frac{1}
{|\vec{x}|} \right ]
\label{iden}
\ee

 The $\mu -$operator is constructed in the following way
\be
\mu (x)=\lim_{\rho ,\delta \rightarrow 0}
\exp \left\{-\fr{i}{6}\int d^{4}z \tilde{A}_{\mu \nu \alpha}(z;x)
(-\Box)^{-1}W^{\mu \nu \alpha}\right\}
\label{mu}
\ee
where $\tilde{A}_{\mu \nu \alpha}$ is the field intensity tensor of
$\tilde{A}_{\mu \nu}$.
 The operator $\mu $ is in principle nonlocal, depending on the
hypersurface $T_{x}(S)$. Nevertheless, as it happens in the case of the above
mentioned related operators \cite{kinks,emprinc} all the hypersurface
dependence of the $\mu $ correlation functions can be eliminated
by the introduction of a  renormalization factor whose
form is uniquely determined only by the requirement of
hypersurface invariance.
The parameters $\rho$ and $\delta$ will be used as regulators which
will be eliminated at the end of the calculations.

 Inserting (~\ref{ext}) in (~\ref{mu}) and observing that the
surface element in (~\ref{ext}) only possesses the 0-component
nonvanishing we get, after integration over $z$
\be
\mu(x) =\lim_{\rho ,\delta \rightarrow 0}\exp\left \{\fr{ib}{
4\pi}\int _{T_{x}(S)}d^{3}
 \vec{\xi}\  \partial _{i}\Phi _{j}( \vec{\xi}- \vec{x})
(-\Box)^{-1}\ W^{0ij}(x^{0},\vec{\xi})\right\}
\label{muw}
\ee

  We can also reexpress $\mu$ in terms of the electromagnetic field
by using
  eq. (~\ref{wa1}):
\be
\mu(x) =\lim_{\rho ,\delta \rightarrow 0}
\exp\left \{ \fr{ib}{4\pi} \int _{T_x(S)} d^3 \vec\xi \
\partial_i\Phi_j (\vec\xi -\vec x) \epsilon^{ijk} A_k (x^0,\vec \xi)
\right\}
\label{mua}
\ee
The charge and topological charge densities,
which are identified by
   (~\ref{wa}) are given, respectively, by
\be
\rho(x)=\partial _{i}E^{i}=\frac{1}{2}\epsilon ^{ijk}\partial
 _{i}W_{jk}
\label{ro}
\ee
Using a covariant (Lorentz) gauge quantization ( see
\cite{ty} for instance) we get the equal-time commutation relations
$[A^i(x),E_j(y)] = i \delta^i\ _j\delta^3(x-y)$ where $E^i =F^{i0}$
is the electric field.

 The commutator between $\mu $ and the charge density can be obtained
by writing $\mu \equiv     e^A$,  using the expansion for
 $e^{A}Be^{-A}$,
Eq.(~\ref{mua}) and the
 commutator $[A^i,E_j]$. The result is
\be
[\rho(y),\mu (x)]=\fr{b}{4\pi}\mu (x)\lim_{\rho ,\delta\rightarrow 0}
\int_{T_{x}(S)}d^{3}\vec{\xi}\ \epsilon^{ijk}\partial^{(y)}_{k}
\delta^{3}(\vec{\xi}-\vec{y})\partial^{(\xi)}_{i}\Phi_{j}(\vec{\xi}-\vec{x})
\label{comro}
\ee
where $\rho (x)$ stands either for the charge or topological charge density.

 Using the identity
 $\del^{(y)}_i= -\del^{(\xi)}_i$ and the results of Appendix D
we can immediately evaluate (\ref{comro}) to be

\be
[\rho (y),\mu (x)]_{ET}=b\mu (x) \delta^3(\vec x -\vec y)
\label{comro1}
\ee
 This equation shows that the operator $\mu$ does in fact carry
 $b$ units of charge.

 Another interesting commutator involving the operator $\mu$ is the one
with the field $W_{\mu\nu}$. From (\ref{wa}), we can express
This field in terms of the electromagnetic field. In the
Lorentz gauge ($\del_\mu W^{\mu\nu}=0$), we have
\be
W^{\mu\nu} = \fr{1}{2} \epsilon^{\mu\nu\alpha\beta} F_{\alpha\beta}
\label{wf}
\ee
{}From this and (\ref{mua}) we can obtain, along the same lines as we
did for the charge commutator, the equal times relation
$$
\mu(x) W_{ij}(y) = \left [ W_{ij}(y) - \fr{b}{4\pi}
 \epsilon_{ijk} \fr{(\vec y -
\vec x)^k}{|\vec y -\vec x|^3} \right ] \mu (x)
$$
\be
  \mu(x) W_{i0}(y) = W_{i0}(y) \mu(x)
\label{comw}
\ee

 This shows that  the operator $\mu$ has indeed the same
kind of commutation relation with the lagrangian field as other
topological excitation creation operators as the vortex \cite{vortices}
and magnetic monopole  operators, for instance \cite{mon}. Observe
that through commutation with $W_{\mu\nu}$, the $\mu$ operator
introduces a configuration having a nonzero topological charge.

 Another interesting relation which can be obtained very in the
same way as (\ref{comw}) is the commutator of $\mu$ with the
electric field:
\be
[E^{k}(y),\mu (x)]_{ET}=b\frac{(\vec{y}-\vec{x})^{k}}{4\pi |\vec{y}
-\vec{x}|^{3}}\ \  \mu (x)
\label{come}
\ee
 In Appendix A it is shown that  this implies
 \be
\frac{<\mu (x)|E^{k}(y) |\mu (x)>_{ET}}{<\mu(x)|\mu(x)>}
=b\frac{(\vec{y}-\vec{x})^{k}}{4\pi |\vec{y}
-\vec{x}|^{3}}
\label{vev}
\ee
The vacuum expectation value of the electric field in
the
 states created by $\mu$ is the
field configuration of a point electric charge of magnitude b.
This characterizes $|\mu (x)>$ as a coherent state
of photons. Observe that the nonvanishing divergence of the
above equation is guaranteed by the r.h.s. of
 (~\ref{newmax1}) even in the absence of matter.
This also shows that the states created by $\mu$ for which
the charge $b$ is different from the charge $q$
do not belong to QED in spite of the fact that we can express
the $\mu$-operator in terms of $A_\mu$. They belong to the additional
nontrivial sectors of the $W_{\mu\nu}$- theory.

 Let us remark at this point the similarity with the two-
dimensional case studied in Section 2. Even though no polynomial
in  the lagrangian
fields can create states carrying the topological charge
 we have an operator -  $\mu$ in our case, $\sigma$ in the
two-dimensional example - expressed nonperturbatively in terms
of the lagrangian fields which do indeed create states bearing
this charge. In the next Section, we are going to evaluate
the two-point correlation function of $\mu$.

\section{ The $\mu$ Correlation Functions }

\setcounter{equation}{0}

 Let us compute in this section the euclidean correlation functions
of the operator  $\mu$ introduced above.
We will consider the case in which matter is not present.
Taking the
expression of $\mu$, Eq. (~\ref{mu}), the lagrangian ${\cal L}_W$
  and going to
euclidean space (in which we will work henceforth) we may write
$$
<\mu (x)\mu ^{\dagger}(y)>=Z^{-1}
\int DW_{\mu \nu} \exp\left \{-\int d^{4}z \left [\fr{1}{12}
W^{\mu\nu\alpha}(-\Box^{-1} W_{\mu\nu\alpha} + \right. \right.
$$

\be
\left. \left.
+\fr{1}{6}\tilde{A}_{\mu \nu \alpha}(z;x,y)(-\Box)^{-1}
W^{\mu \nu \alpha} +
{\cal L}_{GF}+{\cal L}_{CT}
 \right ]  \right \}
\label{cf1}
\ee
where ${\cal L}_{GF}$ is the gauge
fixing term given by (~\ref{gf}) and
 $\tilde{A}_{\mu \nu \alpha}
(z;x,y)=\tilde{A}_{\mu \nu \alpha}(z;x)-\tilde{A}_{\mu \nu \alpha}(z;y)$,
 the minus sign of the y-factor corresponding to the fact that we
have $\mu^\dagger(y)$.
  ${\cal L}_{CT}$ is the above
mentioned  hypersurface renormalization factor to be determined
below.

We see that $<\mu\mu^{\dagger}>=e^{F[\tilde{A}_{\mu \nu}]}$ is the vacuum
functional in the presence of the external field $\tilde{A}_{\mu \nu}$.
This property of the correlation functions of $\mu$
 is common to all of the above mentioned topological charge bearing
related operators
\cite{kinks,vortices,mon} and
follows from the general fact that  topological charge carrying
operators are closely related to the
 disorder variables of Statistical Mechanics \cite{soldis}. Indeed,
treating these operators as disorder variables
 one can demonstrate in general \cite{evora} that the
 $\mu$ operator correlation functions can be expressed in terms of
the coupling of the lagrangian field to an external field like
$\tilde{A}_{\mu \nu }$ as in (~\ref{mu}). One can also show in general
\cite{evora}
that the appropriate
renormalization factor consists of the
corresponding self-coupling of the
the external field.
 Also here, we will see explicitly that the renormalization
counterterm
\be
{\cal L}_{CT}=\fr{1}{12}\tilde{A}^{\mu \nu \alpha}(-\Box)^{-1}
\tilde{A}_{\mu \nu \alpha}
\label{lct}
\ee
will absorb all the hypersurface dependence of the correlation
function, thereby making it completely local.
 We can understand the reason for this in the following way.
Neglecting the gauge fixing term for a while,
 we see that we can write
\be
<\mu(x)\mu^\dagger(y)> = e^{F[\tilde A_{\mu\nu} (T_x(S);x,y)]}=
Z^{-1} \int D W_{\mu\nu} \exp \left\{ W_{\mu\nu} + \tilde A_{\mu\nu}
(T_x(S);x,y) \right\}
\label{wmaisa}
\ee
 Now, let us make the change of functional integration  variable
 \be
 W_{\mu\nu} \longrightarrow W_{\mu\nu} + \del_\mu \Lambda_\nu -
\del_\nu \Lambda_\mu
\label{cv}
\ee
with
\be
\Lambda_\mu = \Theta_4(\Delta V)[ \Phi_\mu(z-x) -\Phi_\mu(z-y)]
\label{lamb}
\ee
where $\Phi_\mu$ was introduced above and
 $\Theta_4(V)$ is the 4-dimensional Heaviside function with
support on the 4-volume $\Delta V$ which is bounded by the
hypersurface $T_x(S)$ and another arbitrary hypersurface
$\tilde T_x (S)$, both of them limited by the surface $S$.
 It is easy to see that under (\ref{cv}) and (\ref{lamb})
the hypersurface $T_x(S)$ in (\ref{wmaisa}) will be exchanged
by $\tilde T_x(S)$ (actually the general form of the extenal field
$A_\mu(z;x)$ contains a 4-volume term which vanishes whenever we
choose the hypersurface to lie in the flat {\myfont R}$^3$ plane
as we are doing here
\cite{vortices,mon}).
This fact clearly shows that the correlation function (\ref{wmaisa})
becomes completely hypersurface independent with our choice
(\ref{lct}) for the renormalization factor.
 When we take in account the presence of the
gauge fixing term, we can still go through the
same procedure above by just exchanging the
singular gauge transformation (\ref{cv}) by the
corresponding BRST transformation \cite{vortices,mon}.

In what follows, we are
going to see by explicit computation
that the correlation function (\ref{cf1}), with the choice
(\ref{lct}) made for $\cl_{CT}$,
is indeed hypersurface independent.
 Let us remark that in the presence of a matter coupling, in order
to attain hypersurface invariance and therefore locality, by the
arguments just exposed, one would have to add a $j^\mu$ dependent
piece to the operator $\mu$. This, of course should not change the
properties of the this operator studied in Section 4.

 Before performing the functional integration in (\ref{cf1}),
let us observe that we can rewrite the linear term as
\be
\fr{1}{6} \int d^4z \tilde A_{\mu\nu\alpha}(-\Box)^{-1}
W^{\mu\nu\alpha} =\fr{1}{2} \int d^4z B_{\mu\nu} (-\Box)^{-1}
W^{\mu\nu}
\label{lin}
\ee
where
\be
B_{\mu\nu}(z;x) = \fr{b}{4\pi} \int_{T_x(S)} d^3\xi_\alpha
\del_\gamma \Phi_\beta (\xi - x) \delta^4(z- \xi) F^{\alpha\beta
\gamma}\ _{\mu\nu}
\label{be}
\ee
and
\be
F^{\alpha\beta\gamma}\ _{\mu\nu} =\del^\alpha \Delta^{\beta\gamma}
\ _{\mu\nu} + \del^\beta \Delta^{\gamma\alpha}\ _{\mu\nu}
+ \del^\gamma \Delta^{\alpha\beta}\ _{\mu\nu}
\label{efe}
\ee

 Integrating
 over $W_{\mu \nu}$ in (~\ref{cf1}) with the
help of the euclidean propagator of this field, Eq. (~\ref{pro}),
we obtain
$$
<\mu(x)\mu^\dagger(y)> =\lim_{\rho ,\delta \rightarrow 0}
\exp \left \{ \fr{1}{2}\int d^4z d^4z'B^{\mu\nu}(z)(-\Box)^{-1}
B^{\alpha\beta}(z') (-\Box)^{-1} D_{\mu\nu\alpha\beta}(z-z')\right.
$$
\be
\left.
- S_{CT} \right \}
\label{cf3}
\ee
 We immediately see that only the first term of (\ref{pro})
contributes to (\ref{cf3}). In particular all the gauge dependence
disappears.
 This happens
because of the gauge invariant way in which the external field is
coupled in (~\ref{cf1}). Using the identity
\be
F^{\mu\nu\alpha}\ _{\sigma\tau} F^{\gamma\rho\beta}\
_{\lambda\chi} = -4 \epsilon^{\mu\nu\alpha\sigma}
\epsilon^{\gamma\rho\beta\lambda} [ -\Box \delta^{\sigma\lambda}
+ \del^\sigma \del^\lambda ]
\label{identidade}
\ee
and performing the $z$ and $z'$ integrals in (\ref{cf3}), we get
$$
<\mu (x) \mu ^{\dagger}(y)>=\lim_{\rho ,\delta ,m,\epsilon
\rightarrow 0}\exp \left \{
 \fr{b^{2}}{32\pi^2}\sum_{i,j=1}^{2}\lambda_{i}\lambda_{j}
\int_{T_{x_{i}}}d^{3}\xi_{\mu}\partial_{\alpha}^{(\xi)}\Phi_{\nu}
(\vec{\xi}-\vec{x_{i}}) \right.
$$

$$
\times\int_{T_{x_{j}}}d^{3}\eta_{\gamma}\partial_{\beta}
^{(\eta)}\Phi_{\rho}(\vec{\eta}-\vec{x_{j}})\ \ \epsilon^{\mu \nu \alpha
 \sigma}\epsilon^{\gamma \rho \beta \lambda}
$$

\be
\left.
\times [ -\Box\delta_{\sigma
\lambda}+\partial^{(\xi)}_{\sigma}\partial^{(\xi)}_{\lambda}]\left
[-\fr{1}{8\pi^{2}}\ln\ m\gamma [\ |\xi -\eta|+|\epsilon |\ ]\ \right ]-S_{CT}
\right \}
\label{cf4}
\ee
In this expression, $x_{1}\equiv x$ , $x_{2}\equiv y$ , $\lambda_{1}
\equiv +1$ and $\lambda_{2} \equiv -1$. The last expression between
brackets is the inverse Fourier transform of $1/k^4$. It
comes from the $(-\Box)^{-1}$ terms of (~\ref{muw}),
 (~\ref{cf1}) and (\ref{cf3}).
 In (\ref{cf4}), $m$ is an infrared regulator used to control the
the small $k$ divergences of the inverse Fourier transform of $1/k^4$
and $\gamma$ is the Euler constant.
 We also introduced the
ultraviolet cutoff $|\epsilon |$ in order to control the
short distance singularities of ${\cal F}^{-1}[1/k^4]$.

 The r.h.s. in (\ref{cf4}) contains two terms proportional to
$\Delta^{\sigma\lambda}$ and $\del^\sigma \del^\lambda$, respectively.
 We are going to see that the second one is hypersurface independent
and leads to a local correlation function. The first one is
hypersurface dependent. We show in Appendix B that it is identical to
$S_{CT}$ and therefore is exactly canceled.

 Let us make now an important observation concerning the renormalization
counterterm $S_{CT}$ which as we said is identical to the
first term in (\ref{cf4}). This contains two pieces:
 the crossed terms (with $i\neq j$)
 and the self-interaction terms ( with $i=j$).
In Appendix C we show that the crossed terms
 vanish.
 The self-interaction terms,  on the
other hand, diverge in this limit. We conclude therefore that the
renormalization counterterm $S_{CT}$ contains {\em only the
unphysical  self interaction
terms}. If the crossed terms were nonvanishing our renormalization
procedure would be meaningless since it would be removing
a nontrivial interaction.

 Let us consider now the second term in (\ref{cf4}). The
derivatives $\del_\sigma^{(\xi)}$ and $\del_\lambda^{(\xi)} =
-\del_\lambda^{(\eta)}$ can be made total because of the
$\epsilon^{\mu\nu\alpha\beta}$
 factors. Using  the Gauss theorem we can transform
the hypersurface integrals in surface integrals. Then, using the
results of Appendix D, we immediately get
\be
<\mu(x)\mu^\dagger(y)>=
\lim_{m,\epsilon \rightarrow 0}
\exp \left
 \{\fr{b^2}{8\pi^{2}}[-\ln m\gamma |x-y|+\ln m \gamma |\epsilon |]
\right \}
\label{cf5}
\ee
Note that the $m\gamma$ factors cancel out. In a charge selection rule
violating correlation function (like $<\mu \mu >$, for instance) we
would have the sign of the $\ln |x-y|$  term reversed and the
$m\gamma$ factors would no longer cancel, actually forcing the
correlation function to vanish in the limit $m\rightarrow 0$ and
thereby enforcing the charge selection rule.
The ultraviolet divergence at $|\epsilon | \rightarrow 0$
can be eliminated by a multiplicative renormalization of the field
$\mu$, namely
\be
\mu_{R}(x)=\mu (x)|\epsilon |^{-b^{2}/16\pi^{2}}
\label{ren}
\ee
 Using this we finally get
\be
<\mu_{R}(x)\mu^{\dagger}_{R}(y)>=|x-y|^{-b^{2}/8\pi^{2}}
\label{cf6}
\ee
This is our final expression for the $\mu$ field
euclidean two-point correlation function.  Observe
that it is completely local. It is remarkably
similar to the $\sigma$-operator correlation
function (\ref{2.8}) of the two-dimensional analog system.
{}From the long distance behavior $\lim_
{|x-y|\rightarrow\infty}
<\mu_{R} (x)\mu_{R}^{\dagger}(y)>=0$, we can infer that
 $<\mu_R (x)>=0$ and therefore
that the states  $|\mu (x)>$ , created by $\mu$ are
orthogonal to the vacuum and
therefore nontrivial. The power law decay of the  correlation
function on the other hand implies that these states
are massless.

 The form of the correlation function (\ref{cf6}) implies that
the norm of the states created by $\mu$ is positive definite.
 Indeed, going back to the Minkowski space, we have
$$
\| |\mu_R> \|^2\equiv \lim_{x \rightarrow y ; (x-y)^2 < 0}
<\mu_R (x)| \mu_R (y)> =
$$
\be
 \lim_{x \rightarrow y ; (x-y)^2 < 0}
\left [ |\vec x -\vec y|^2 - (x^0 - y^0)^2 \right ]^{-b^2/8\pi^2} > 0
\label{norm}
\ee
 This property shows that  the states created by $\mu$ belong to the
physical sector of the Hilbert space
of the $W_{\mu\nu}$ theory
even though they do not belong to QED which,
in the absence of matter, constitutes the zero topological
 charge sector.
 Here again we find the
analogy with the 1+1 D massless scalar field where the corresponding
above mentioned $\sigma$ operator  creates states of positive norm
in spite of the fact that the Hilbert space has indefinite metric.

Observe that
the renormalization process used to obtain local correlation
functions starting from the bare $\mu$ of course does not change the
commutation rules of this field. We would like to stress that although
the expression (~\ref{mua}) of $\mu$ in terms of the electromagnetic
field is well suited for the obtainment of commutation rules,
the natural form of $\mu$ which combines with the renormalization
counterterm ${\cal L}_{CT}$ and the lagrangian ${\cal L}_W$ in
order to to produce local correlation functions is the one
expressed in terms of $W_{\mu\nu}$, Eq. (~\ref{muw}).

 An arbitrary 2n-point correlation
function could be obtained in a straightforward manner by just
inserting additional external fields $\tilde{A}_{\mu \nu}$ in
an expression like (~\ref{cf1}). They would be
given by products of monomials of the type we found in
$<\mu\mu^\dagger>$.

 In the presence of matter
 we would no longer be able to obtain an exact
expression for the $\mu$ correlation functions because the
integration over the matter fields could no longer be done exactly.
Nevertheless, due to the well known infrared asymptotic freedom
of QED we still may expect that the long distance behavior of
$<\mu \mu^{\dagger}>$ will be the same. The result
that the operator $\mu (x)$ creates massless states orthogonal to the
vacuum, therefore, also holds in the presence of matter fields.

   Before finishing this section, let us mention
a well known operator which shares some of the properties
of $\mu$ as the commutator with the charge operator,
for instance. This is
  the Wilson line,
 $\exp \{-i(b/4\pi)$ $\int^{x}A_{\mu}dx^{\mu}\}$. This operator
is not appropriate because
for it we would have
nonvanishing string dependent crossed ($i\neq j$) interaction terms
 (see the remarks after Eq.(~\ref{cf4}) and
Appendix C) which would inevitably
lead to a nonlocal correlation function.

\vfill
\eject

\section{Phenomenological Estimates }

\setcounter{equation}{0}

  In this Section, let us try to infer something about
the observational characteristics of the topological excitations
we have just described. Of course all that follows would only
be
valid in the event the more general description of QED, based on
the antisymmetric tensor gauge field, would in fact be realized
in nature. This is by no means guaranteed by the present formalism.

\subsection{ The Quantum of Charge }

The charge of the states created
 by $\mu $ is in principle not quantized since the
parameter
 $b$,
which determines the value of the charge of the states created by
$\mu$
 is arbitrary, in the same way as in the two-dimensional
 analogous system.
There is, however,  an important difference
from the 1+1 D case. Considering the spectral representation of the
$\mu$ correlation function
\be
<\mu_{R}(x)\mu^{\dagger}_{R}(y)>=|x-y|^{-b^{2}/8\pi^{2}}=
\int_0^\infty dM^2 \rho(M^2) \int \fr{d^4k}{(2\pi)^4}
\fr{e^{ik\cdot(x-y)}}{k^2 +M^2}
\label{sr}
\ee
we immediately see that the spectral density is of the form
\be
 \rho(M^2) = \left\{ \begin{array}{ll}
  \lambda \  (M^2)^\sigma  & \mbox{$b\neq 4\pi$} \\
  \lambda ' \  \delta(M^2)   & \mbox{$b=4\pi$}
                      \end{array}
                    \right.
\label{sd}
\ee
where $\lambda$, $\lambda '$ and $\sigma$ are constants.

 The spectral density is in
general a power function except for $b=4\pi$ when it is a delta
function with support on a mass equal to zero. We therefore conclude
that only for this value of $b$ the $\mu$-field
 really interpolates asymptotic states.
  We see that in this case, the
$\mu$- correlation function behaves at long distances
  as a free massless scalar field correlation function.
In the above mentioned 1+1 D system, on the other
hand, the spectral density of the
$\sigma$-field correlation function has no delta function
singularity for any value of $b$
 and therefore the asymptotic states associated with
$\sigma$ are absent from the spectrum.

 Based on the above considerations and imposing the
condition that the charged topological states should
be observable asymptotically, we can fix their charge to be
 $b=4\pi$. This, of course, is in the natural units system
where the fine structure constant is $e^2/4\pi = 1/137$ and
the electron charge is $e=(4\pi/137)^{1/2}$. The charge of
these states, therefore can be expressed in terms of
the electric charge as $b=(4\pi 137)^{1/2}e$.

It would be interesting to investigate
whether some additional coupling
to the $W_{\mu\nu}$ field $-$ as it happens with the
Sine-Gordon coupling in the case of the scalar field in 1+1 D $-$
would produce a charge quantization for the topologically
charged states.
This would have interesting consequences in the
presence of matter where some insight could possibly
be obtained on the quantization of matter charge since, as we saw,
in this case there are sectors where the topological charge
is identified with the matter charge.

\subsection{ The Intensity Threshold }

  Let us show here that the
eventual observation of the charged topological states of the
gauge field would require the presence of
 an extremely intense backgroud
electromagnetic field. We can understand this by examining the
expression (\ref{mua}) of $\mu$ in terms of the electromagnetic
field and the correlation function (\ref{cf6}). First of all, let
us remark that we can add in (\ref{cf5}) a $\ln \ell_0$ to
both terms in the exponent. This will fix the scale
of the large distance regime
of the correlation function.
 We can therefore write and expand the $\mu$  correlation function (\ref{cf6})
as
$$
<\mu_{R}(x)\mu^{\dagger}_{R}(y)>=\left [
\fr{|x-y|}{\ell_0}\right ]^{-b^{2}/8\pi^{2}} =
\exp \left [ -\fr{b^2}{8\pi^2} \ln \fr{|x-y|}{\ell_0} \right ]=
 $$
\be
 \sum_{n=0}^\infty (-1)^n  \fr{
 \left [ \fr{b^2}{8\pi^2} \ln \fr{|x-y|}{\ell_0} \right ]^n}
{n!}
\label{cfexp}
\ee

 If we look at expression (\ref{mua}) for $\mu$ and
expand the exponential which appears there, we immediately
realize that the order $n$ in the above expansion of the
$\mu$ correlation function is nothing but the number of photons
contributing to it. Hence, we can discover the minimal
number of photons which is necessary to build up the coherent
topological states by determining until what order $N$ the
terms in the expansion (\ref{cfexp}) give an important
contribution.

 Observe now that in the presence of matter $-$ which is always
the real situation, because at least the vacuum fluctuations of matter
we can never remove $-$ expression  (\ref{cfexp}) represents the
large distance regime of the $\mu$ correlation function. The
only scale which fixes the magnitude of large distances is the
electron Compton wavelegth $\lambda_e$ and therefore we make
the identification $\ell_0 \equiv \lambda_e$ in (\ref{cfexp}).

 Using the value $b=4\pi$ which we found above and choosing
\be
\fr{|x-y|}{\lambda_e} \sim 10
\label{sca}
\ee
for the onset of the large distance regime we find that
for $n \simeq 14$ the summand in (\ref{cfexp}) is of the
order of $0.02$ and decreases fastly for higher values of $n$.
 We conclude, therefore, that $N\simeq 14$ is the order of
magnitude of the minimal number of photons needed to produce
the topological states. These photons, however, must be
within a region of linear dimension of the order of $10 \lambda_e$,
according to our choice (\ref{sca}).

   The intensity of the background field is now readily estimated
by using the formula
\be
 I=\fr{N E c}{V}
\label{intens}
\ee
where $E$ is the photon energy, $c$ is the speed of light and
$V$ is the volume where the minimal number of photons should be.
 As we argued above, $V\simeq 10^3 \lambda_e^3$ and $N\simeq 14$.
 Taking the photon wavelenght to be of the order $\lambda \simeq
1 \mu m$ \cite{km}
we get
\be
I \simeq 6 \times 10^{18} {\rm watt}/{\rm cm}^2
\label{int1}
\ee
 To have an idea of how large this intensity is, we compare it
to the intensity of the electromagnetic radiation at the surface of the
sun which is of the order $I_\odot \simeq 6\times 10^3 {\rm watt}/
{\rm cm}^2$. The fact that the threshold intensity is so high would
preclude the production of the $\mu$-states in ordinary processes
like electron-positron annihilation, for instance. Laser fields
of intensities of the order of the threshold we estimated here,
however, can be constructed \cite{km}. There is actually a projected
experiment at SLAC in which high energy electrons are scattered
by such a laser field \cite{km}. This is precisely the kind of
situation in which the charged topological excitations we studied
here should be observed if it is indeed true that the formulation
we introduced here is realized in nature.

\section{ Conclusion}

  We have seen that it is possible to introduce
a new formulation of QED  in terms of
an antisymmetric tensor gauge field whose
 topological current
appears
as a source for the electromagnetic field.
 The charged states of QED lie in the sector where the
topological current is identical to the charged matter current.
 There are, however, new nontrivial topologically charged
sectors whose charge is not associated to matter but
only to the gauge field. This charge would interact
with the electromagnetic field and with the  charged matter
states of QED in the usual way described by QED.
 The topologically charged
states are created by a gauge invariant operator and the long distance
behavior of the correlation functions of this operator indicates that
these states are massless. Imposition of the existence of the
asymptotic states interpolated by the topological excitation
creation operator fixes the value of their charge.

 Of course it may well be that this more general formulation of
QED is just not realized in nature. If, on the other hand,
should it turn out to provide a correct description $-$
  in which event the topologically charged states
 would be possibly  observed $-$ an extremely intense
background electromagnetic field ($I\simeq 10^{18} {\rm watt}/
{\rm cm}^2$) would be needed in order to produce them.
  Laser fields of this order of intensity can be obtained \cite{km}.
Scattering in the presence of such fields would be the ideal
framework to test the validity of the formulation introduced here.
  We are presently computing the cross section for the photoproduction
of a $\mu^+ \mu^-$ in the presence of such strong background fields
\cite{cs}.

 There are some interesting questions which arise in connection
with these new topologically charged states: is there any
mechanism of mass generation for them ? What happens when we
couple other fields like in the Electroweak theory, for instance ?
What about the usual sectors where the toplogical charge
is identical to the matter charge ? Could this formulation
shed some light on the problem of quantization of matter charge ?

 The results we found in this work indicate that
electric charge, which is a physical quantity
 usually associated with matter can be
 obtained as an attribute of some
coherent states of the gauge field itself. It is not
inconceivable that other quantities like spin, mass, flavor, color
and so on could be generated as well as properties of some
peculiar states of the gauge fields in general. This would lead to the
outstanding possibility of describing both matter and the
fields which mediate its interactions within the same unified
framework. We are sure this possibility is very far from what
has been presented here but we hope it could be a step
towards this end.

\vskip 10mm

\leftline{\Large\bf Acknowledgements}

\bigskip

I would like to thank the Physics Department of Princeton University
and especially C.Callan and D.Gross for the kind hospitality.
 I have benefited from many stimulating and enlightening
discussions with several people. I am especially
 grateful to C.Callan, C.Teitelboim, K.McDonald, A.Polyakov,
S.MacDowell, I.Kogan and S.Dalley.
 I am also grateful to the Brazilian National Research Council
(CNPq) for financial support.

\vfill
\eject

\appendix

\renewcommand{\thesection}{\Alph{section})}

\renewcommand{\theequation}{\Alph{section}.\arabic{equation}}

\setcounter{equation}{0}

\section{ Appendix A}

Let us demonstrate here eq. (\ref{vev}). Starting from (\ref{come})
and applying it on $<\mu(x)|$ to the left and on the vacuum to the
right, we get
\be
<\mu (x)|E^{k}(y) |\mu (x)>_{ET} =
<\mu(x)| \mu(x) E^k(y) |0> +
b\frac{(\vec{y}-\vec{x})^{k}}{4\pi |\vec{y}
-\vec{x}|^{3}} <\mu(x)|\mu(x)>
\label{A.1}
\ee
Since
\be
\lim_{x\rightarrow y} [\mu^\dagger(x)\mu(y), E^k(z)] =0
\label{A.2}
\ee
it follows that the second term in (A.1) vanishes and
(\ref{vev}) is immediately established.

\section{Appendix B}

\setcounter{equation}{0}

 Let us demonstrate here that the first term of (\ref{cf4})
is identical to $S_{CT}$.
 To begin with, let us show that
\be
I=
\int d^4z \del_\alpha A_{\mu\nu}(z;x) T^{\alpha\mu\nu}(z)
=
\fr{b}{4\pi} \int _{T_x(S)} d^3\xi_{[\mu} \del^{(\xi)}
_\alpha \Phi_{\nu]} (\xi -x) T^{\alpha\mu\nu} (\xi)
\label{B.1}
\ee
 where $A_{\mu\nu}$ is given by (\ref{ext})
and $T^{\mu\nu\alpha}(z)$ is an arbitrary tensor.
According to (\ref{ext})
we have
\be
\del_\alpha^{(z)} A_{\mu\nu}(z;x) = \fr{b}{4\pi}
\int_{T_x(S)} d^3\xi_\mu \Phi_\nu(\xi -x) \del_\alpha^{(z)}
\delta^4(z- \xi) + \fr{b}{4\pi} \hat n^\mu \oint _S
d^2\xi_\alpha  \Phi_\nu(\xi - x) \delta^4 (z-\xi) -(\mu
\leftrightarrow \nu)
\label{B.2}
\ee
where $\hat n^\mu$ is the unit vector in the direction of
$d^3\xi ^\mu$.
  Using Gauss' theorem in the last term and integrating in $d^4z$
with $T^{\alpha\mu\nu}$ we get
$$
I=
\fr{b}{4\pi} \int _{T_x(S)} d^3\xi_{[\mu}
 \Phi_{\nu]} (\xi -x) (-\del^{(\xi)}_\alpha)
T^{\alpha\mu\nu} (\xi) +
\fr{b}{4\pi} \int _{T_x(S)} d^3\xi_{[\mu} \del^{(\xi)}
_\alpha \left [ \Phi_{\nu]} (\xi -x) T^{\alpha\mu\nu} (\xi)
\right ]
$$
\be
\fr{b}{4\pi} \int _{T_x(S)} d^3\xi_{[\mu} \del^{(\xi)}
_\alpha \Phi_{\nu]} (\xi -x) T^{\alpha\mu\nu} (\xi)
\label{B.3}
\ee
 thus establishing (\ref{B.1}).
 Now consider the first term of (\ref{cf4}). Antissymmetrizing in
$\mu ,\nu$ and  $\gamma , \rho$ and using the fact that the
last expression between brackets is $(-\Box)^{-2}$ as well as
(B.3), we can write the first term of (\ref{cf4}) as
$$
\fr{1}{8} \int d^4zd^4z' \epsilon ^{\mu\nu\alpha\sigma}
\del_\alpha A_{\mu\nu}(z)\left [ \fr{1}{-\Box} \right ]
\epsilon^{\gamma\rho\beta\sigma} \del '_\beta A_{\gamma\rho}=
$$
\be
\fr{1}{12} \int d^4z A_{\alpha\mu\nu}(-\Box)^{-1}
A^{\alpha\mu\nu}
\label{B.4}
\ee
which is precisely $S_{CT}$.

\section{ Appendix C}

\setcounter{equation}{0}

 Let us show here that the $i \neq j$ terms of $S_{CT}$ vanish in
the limit when $\rho \rightarrow 0$.
 From (\ref{cf4}), we see that these terms are proportional to
$$
I=
\int_{T_{x_{i}}}d^{3}\xi_{\mu}\partial_{\alpha}^{(\xi)}\Phi_{\nu}
(\vec{\xi}-\vec{x_{i}})
$$

\be
\times\int_{T_{x_{j}}}d^{3}\eta_{\gamma}\partial_{\beta}
^{(\eta)}\Phi_{\rho}(\vec{\eta}-\vec{x_{j}})\ \ \epsilon^{\mu \nu \alpha
 \lambda}\epsilon^{\gamma \rho \beta \lambda}
  \cf^{-1}[ (-\Box)^{-1}]
\label{C.1}
\ee

Using the fact that $d^3\xi^\mu$ //  $d^3\eta^\gamma$ // $
\hat n^4$ and the identity (\ref{iden}), we can write (C.1) as
\be
I=
\int_{T_{x_{i}}}d^{3}\xi_{\mu}\partial_{\alpha}^{(\xi)}
\left [ \fr{1}{
|\vec{\xi}-\vec x|} \right ]
\int_{T_{x_{j}}}d^{3}\eta^{\mu}\partial^{\alpha}
_{(\eta)}\left [ \fr{1}{|\vec{\eta}-\vec y|}\right ]
  \cf^{-1}[ (-\Box)^{-1}]
\label{C.2}
\ee
Taking the same steps which led us to (B.1), we can write
\be
I= \int d^4z d^4z' \del^\alpha C^\mu (z;x) [ (-\Box)^{-1} ]
\del'_\alpha C_\mu(z';y)
\label{C.3}
\ee
where
\be
C^\mu(z;x) = \int _{T_x(S)} d^3\xi ^\mu \left [\fr{1}{
|\vec \xi - \vec x|}\right ] \delta^4 (z-\xi)
\label{C.4}
\ee
Integrating by parts the derivatives in (C.3), we get a
delta function. Subsequent integration over $z'$ gives
$$
I= \int C^\mu(z;x) C_\mu(z;y)=
$$
\be
\int_{T_x(S)} d^3\xi^\mu \int_{T_y(S)} d^3\eta^\mu
\fr{1}{|\vec \xi -\vec x|} \delta^4(\xi -\eta)
\fr{1}{|\vec \eta - \vec y|}
\label{C.5}
 \ee

  This expression vanishes due to the
presence  of the delta function and because of the fact that
the hypersurfaces $T_x$ and $T_y$ can
always be chosen so as to have an empty intersection.

 For the $i=j$ terms we would have $T_x =T_y$ and the analogous
integral would be seen to diverge.

\section{ Appendix D }

\setcounter{equation}{0}

 Let us demonstrate here how to  obtain results (\ref{comro1}) and
(\ref{cf5}). In both cases, we have  one or
more integrals of the type
\be
\lim_{\rho ,\delta \rightarrow 0}
\int _{T_x(S)} d^3\xi \epsilon^{ijk} \del_i^{(\xi)} \Phi_j
(\vec \xi - \vec x) \del_k^{(\xi)} F(\xi)
\label{D.1}
\ee
for some function $F(\xi)$.
 The derivative $\del_k$ can be made total because of the rotational.
Then, we can use Gauss' theorem to get
\be
\lim_{\rho ,\delta \rightarrow 0}
\oint _{S(x)} d^2 \xi ^k
 \epsilon^{ijk} \del_i^{(\xi)} \Phi_j
(\vec \xi - \vec x)  F(\xi)
\label{D.2}
\ee
 Now, making use of the identity (\ref{iden}), we see that the
cone piece of $S(x)$ does not contribute to (D.2) and we can
already take the limit $\delta \rightarrow 0$.
  Inserting (\ref{iden}) in (D.2) we get
\be
\lim_{\rho \rightarrow 0}
\oint _{{\rm Sphere}(x)} d \Omega \rho^2 \left [\fr{1}{\rho^2}
\right ] F(\xi)= 4\pi F(x)
\label{D.3}
\ee

This result leads immediately to (\ref{comro1}) and (\ref{cf5}).

\vfill
\eject

\centerline{\Large\bf Figure Caption}

\bigskip

Fig. 1 - Three dimensional hypersurface used in the definition
of the external field $A_{\mu\nu}(z;x)$ and the operator $\mu$
 (actually a cut of it).


\vskip 5.0mm

\begin{thebibliography}{99}

\bibitem{col} S.Coleman, {\it ``Aspects of Symmetry''}, Cambridge, (1985).

\bibitem{kob} R.K\"{o}berle and E.C.Marino, {\it Phys. Lett.} {\bf 126B  }
 (1983) 475.

\bibitem{W} A.S.Wigthman, {\it ``Introduction to Some Aspects of the
Relativistic Dynamics of Quantum Fields''} in Carg\` ese Lectures
in Theoretical Physics 1964, M.L\'evy editor, Gordon and Breach,
NY (1967).

\bibitem{ama} R.L.P.G. Amaral and E.C.Marino, {\it J. of Phys.}
{\bf A25} (1992) 5183.

\bibitem{ty} D.M.Gitman and I.V.Tyutin, {\it ``Quantization of Fields
with Constraints"}
Springer-Verlag, Berlin (1990).

\bibitem{kinks} E.C.Marino and J.A.Swieca, {\it Nucl.Phys.} {\bf
B170} [FS1], 175 (1980).
E.C.Marino, B.Schroer and J.A.Swieca, {\it Nucl.Phys.} {\bf B200} [FS4],
473 (1982).

\bibitem{vortices}
E.C.Marino, {\it Phys.Rev.} {\bf D38}, 3194 (1988).

\bibitem{mon}
E.C.Marino and J.E.Stephany Ruiz, {\it Phys.Rev.}
{\bf D39}, 3690 (1989)

\bibitem{emprinc} E.C.Marino, {\it ``Duality, Quantum Vortices and
Anyons in Maxwell-Chern-Simons-Higgs Theories"}, Princeton University
report PUPT-1330, {\it Annals of Physics} (1993), in press.

\bibitem{soldis} L.P.Kadanoff and H.Ceva, {\it Phys.Rev.} {\bf B3},
 3918 (1971).
E.Fradkin and L.Susskind, {\it Phys.Rev.} {\bf D17}, 2637 (1978).
J.B.Kogut, {\it Rev.Mod.Phys.} {\bf 51}, 659 (1979).

\bibitem{evora} E.C.Marino, {\it ``Dual Quantization of Solitons"}
in Proceedings of the NATO Advanced Study Institute {\it ``Applications
of Statistical and Field Theory Methods to Condensed Matter"},
D.Baeriswyl, A.Bishop and J.Carmelo , editors (Plenum,New York)(1990)

\bibitem{km} K.T.McDonald, private communication.

 \bibitem{cs} E.C.Marino, {\it ``The Cross Section for
Photoproduction of Topologically Charged Gauge Field States
Associated with QED''}, to appear.


\end{thebibliography}
\end{document}